\documentstyle[prd,preprint,aps,epsfig]{revtex}
\begin{document}

\draft
\def\be{\begin{equation}}
\def\ee{\end{equation}}
\def\bea{\begin{eqnarray}}
\def\eea{\end{eqnarray}}
\def\nn{\nonumber}
\def\ep{\epsilon}
\def\c{\cite}
\def\m{\mu}
\def\ga{\gamma}
\def\lan{\langle}
\def\ran{\rangle}

\def\Ga{\Gamma}
\def\la{\lambda}
\def\si{\sigma}
\def\al{\alpha}
\def\pa{\partial}
\def\de{\delta}
\def\De{\Delta}
\def\rsr{{r_{s}\over r}}
\def\rrs{{r\over r_{s}}}   
\def\rs2r{{r_{s}\over 2r}}
\def\l2r2{{l^{2}\over r^{2}}}
\def\rsa{{r_{s}\over a}}
\def\rsb{{r_{s}\over b}}
\def\rsro{{r_{s}\over r_{o}}}
\def\rss{r_{s}}
\def\a2{{l^{2}\over a^{2}}}
\def\b2{{l^{2}\over b^{2}}}
\def\op{\oplus}
\def\sn{\stackrel{\circ}{n}}
\def\c{\cite}

\title{The general  treatment of  high/low energy particle \\ 
interference  phase  in a gravitational field}

\author{  C. M. Zhang and A. Beesham}
\vskip 0.5cm
\address{Instituto de F\'{\i}sica Te\'orica\\
Universidade Estadual Paulista\\
Rua Pamplona 145\\
01405-900\, S\~ao Paulo \\
Brazil\\
zhangcm@ift.unesp.br}
\maketitle

\begin{abstract}
The   interference  phase of the high/low  energy particle
 in a gravitational field  is  studied. 
 By means of the complete Schwarzshild tetrads, we again
deal with  the  phase in
terms of the type-I and type-II phases, which correspond to the momentum and
 ``spin connection" of Dirac particle coupling to the curved spacetime
respectively. We find that the type-II phase is in the magnitude of 
the square of the gravitational potential, which can be neglected  in the
weak field.
For the high energy particle (mass neutrinos), we obtain that the  phase 
calculated along the null is equivalent to the half  
phase  along the geodesic in the high energy limit. 
Further, we apply the covariant phase to the thermal neutron
interference, and obtain the consistent interference phase with that
exploited in COW experiment.

\end{abstract}

PACS number(s): 95.30.Sf, 26.30.+k, 14.60.Pq


\section{Introduction}
The neutrino oscillations have  been a hot topic in the high energy
experimental
and theoretical phycics recently~\cite{zub98,bil98}, in particular,
with highly confident atomspheric 
neutrino experiment of Super-Kamiokande to assure the neutrino mass
~\cite{sk}. As a
natural extension of the theoretical consideration, the description of
neutrino oscillation in the flat spacetime should be replaced by that in
the curved
spacetime if the gravitational background is taken into account. In other
words, the physics related to the neutrino oscillation in Minkowski
spacetime with Lorentz invariant will be extended to Riemanian spacetime 
with general coordinate transformation. The pioneer theoretical
considerations  on the gravitationally induced neutrino oscillation were 
proposed by Wudka~\cite{wud91}, and later advocated by
Ahluwalia~\cite{ahl96} who for the first time 
presented the three flavor neutrino oscillations in the weak field
expansion scheme. Further some thoughtful idea on the nongeometrical 
element in the description of the gravitational theory, trigered by the 
gravitationally induced neutrino oscillation clock, was also
provided~\cite{ahl96}.
Moreover the violation of the equivalence principle was also employed to 
account for the significant influence on the MSW effect~\cite{msw} for the
solar neutrinos~\cite{gas89,man}. More recently, from the different angle,  
the gravitational effects on the neutrino oscillations have  been
paid much
attention  by a number of
authors~\cite{ahl96,wud96,gro96,ful96,bha99,for96}, but, unfortunately, 
debates and conflicts occur in the understanding of the gravitationally  
induced 
neutrino oscillations
~\cite{ahl96,gro96,ful96,bha99,for96}. 

In this paper, we will discuss the particle interference phase 
in a gravitational field in a 
unified version, i.e.,  provide a unified  description of the
phases  for both high energy particle (two mass neutrinos) and low energy 
particle (COW thermal neutrons).
 In order to develope the  treatments by using the null
condition to calculate the neutrino relative phase and appling the weak
field condition to calculate COW thermal neutron phase, we employ the
accurate particle world line (geodesic) to calculate both phase
factors, and at last we can obtain the correct neutrino relative phase 
by dividing a factor of 2 and obtain the COW neutron 
interference phase by the convenient
approximated condition. 

The energy condition in the gravitational field to account
for the relative phase of mass neutrinos often makes 
the confusions and even 
conflicts\c{ahl96,ful96,bha99}.  Apparently, we point out that
almost all debates related to 
the neutrino oscillation in curved spacetime originate from 
the inconvenient  use of this   condition. It is noted that 
replacing the null by the geodesic  to calculate  the neutrino
oscillation 
phase will produce a factor of 2  error. 
 However this factor of 2  will 
be automatically deleted when considering the two neutrino arrival time
differnce\c{bha99}. 
 Here, we still follow the plane wave treatment for the
extremely relativistic neutrinos in
the framework of the standard treatment~\cite{for96}, 
otherwise the wave packet treatment will be applied for the general
case~\cite{giu91}.

For the reason of simplicity, we confine our treatment in two generation
neutrinos (electron and muon) and mainly  
in Schwarzschild geometry with radial propagation and nonradial
propagation respectively. 
The purpose of the paper is threefold. First we point out 
the general  treatment of the high (low) energy 
particle  phase in a gravitational
field, include the  type-I and type-II phases.
Second we give the complete description of neutrino phase along the
geodesic, and point out the relation to that by  the null in the high
energy limit. 
 Third we apply the covariant  geodesic phase to the COW thermal
interference  phase, then we can find that the geodesic phase  
can result in 
the correct results for the interference phases  of  both the  two mass
neutrinos and the thermal neutron by the standard treatment.

So the paper is organized as follows.   
In Sec. II,  we discuss Dirac equation and the  treatments of particle
interference phase  in the curved 
spacetime, which includes type-I and type-II phases. In Sec. III and Sec.
IV, we calculate    the
 neutrino oscillation 
phase in Schwarzschild spacetime along the geodesic,
 and compare it with that along the null, 
  in the radial and nonradial directions 
respectively. The application of the unified geodesic phase to the COW
thermal neutron interference is done in Sec. V.  
Furthermore,
 discussions and conclusions involved in the  high energy neutrino
oscillations  and low energy thermal neutron interference in
curved spacetime are given.
We set $G = \hbar = c = 1$ throughout this article.

\section{Dirac equation in a curved spacetime}

Previous to entering our main point, we stress that the semiclassical
approximated Dirac particle does not follow the  geodesic exactly,
but the force aroused by the  spin and the curvature coupling
has a little contribution to the geodesic deviation~\cite{aud81}. So here
we take the
neutrino as a spinless particle to go along the geodesic~\cite{wud91}.
The gravitational effects on the spin incorporated into Dirac
equation  through the ``spin
connection'' $\Gamma_{\mu}$ appearing in the Dirac equation
in curved spacetime \cite{dirac,ana},
which is constructed by means of the variation of the covariant Lagrangian
of the spinor field as,

\be
\left[ \gamma^{a} h^{\mu}_{a} (\partial_{\mu} + \Gamma_{\mu})
        + m\right] \psi = 0.
\label{dirac2}
\ee
In this equation and in the rest of this section, greek indices $\mu,
\nu, \si$
refer to general covariant (Riemanian) coordinates, while the
latin indices
$a,b,c,d$ refer to locally Lorentz (Minkowski)
coordinates. The tetrads $h_{a}{}^{\mu}$ connect these sets of
coordinates by

\be
g_{\mu \nu} = \eta_{a b} \; h^a{}_\mu \; h^b{}_\nu \; ,
\label{gmn}
\ee
the tetrads are  supposed to satisfy the following relation

\be
h^{a}{}_{\mu} \; h_{a}{}^{\nu} = \delta_{\mu}{}^{\nu} \quad
; \quad h^{a}{}_{\mu} \; h_{b}{}^{\mu} =
\delta^{a}{}_{b} \; .
\label{orto}
\ee
The explicit expression for $\Gamma_{\mu}$ can be written in terms of 
the Dirac matrices and tetrads(see ~\cite{ful96})
\be
\Gamma_{\mu} = {1 \over 8}[\gamma^b, \gamma^c] h_{b}{}^{\nu}
        h_{c \nu;\mu}.
\ee
We must first simplify the Dirac matrix product in the spin
connection term. It can be shown that
\be
\gamma^a [\gamma^b, \gamma^c] = 2 \eta^{ab} \gamma^c - 
        2 \eta^{ac} \gamma^b - 2i \epsilon^{dabc} \gamma_5
        \gamma_d,  \label{gammas}
\ee
where $\eta^{ab}$ is the metric in flat space and
$\epsilon^{abcd}$ is the (flat space) totally antisymmetric tensor,
with $\epsilon^{0123}= +1$. With Eq.(\ref{gammas}), the
contribution from the spin connection is arranged as 
\be
\Ga_{\mu} = 
 {1\over 2}v_{\mu}  - {3i \over 4}a_{\mu}\ga_{5}, 
\label{ga2}
\ee
where $v_{\mu}$ is the tetrad vector  and $a_{\mu}$ is the
tetrad axial-vector respectively, defined by

\be
v_{\mu}= T^{\la}{}_{\la\mu},\,\,\, a_{\mu}= {1\over 6}
\ep_{\mu\nu\la\si}T^{\nu\la\si}
\ee
        
\be        
T^\si{}_{\mu \nu} = \Ga^{\si}{}_{\nu \mu} - \Ga^ 
{\si}{}_{\mu\nu} \;\;,\;\;
\Ga^{\si}{}_{\mu\nu} = h_{a}{}^\si \partial_\nu
h^a{}_\mu \; ,
\label{tor}
\ee
If $v_{\mu} \rightarrow 0$, then the expression of Eq.(\ref{ga2}) is
recovered to the form 
obtained by Cardal and Fuller~\cite{ful96}. It is instructive to outline
the properties 
of the spin connection. We can see that the two terms in the spin
connection represent the different aspects of the gravitational coupling 
with the Dirac field. The second term of r.h.s. of Eq.(\ref{ga2}),
proportional to $i\ga_{5}$, has the
similar form to the weak interaction field~\cite{boe92,bah94}. 
 Moreover $a^{\mu}$ is an
axial-vector, which represents the physical modification from axial
symmetry to the spherical symmetry~\cite{nit81}. From the gravitational
interation 
point of view, $a^{\mu}$ shows the rotational gravitational field
property, or the angular momentum aspect of gravitational field, in other
words, it represents the gravitomagnetic like interaction. 
However the first term of Eq.(\ref{ga2})  is similar to 
the canonical momentum, then its contribution seems to be the
gravitational
field momentum.  In order to group Eq.(\ref{ga2}) with terms arising from
the 
matter effects, we can without physical consequence arrange Eq.(\ref{ga2})
as
follows
 
\be
\Gamma_{\mu} =  P_{G\mu} + i A_{G\mu} {\cal P}_L ,
\label{gravpot}
\ee

\be
A^{\mu}_G \equiv {h \over 4}  h_a{}^{\mu} \epsilon^{abcd}
        (h_{b\nu,\sigma} - h_{b\sigma,\nu}) h_c{}^{\nu}   
        h_d{}^{\sigma} = {3h\over 2} a^{\mu}, 
\ee

\be
P_{G\mu}= {1 \over 2}v_{\mu}\;\;, \;\;
{\cal P}_L = \left[-{\gamma_5 \over 2h}\right]. 
\ee
In these equations, $h = (-g)^{1/2} = [\rm det(g_{\mu\nu})]^{1/2}$. 
The expression in Eq.(\ref{gravpot}) treats left- and right-handed
states differently. 
Proceeding as in the discussion of matter effects, we will borrow the
neutrino oscillation standard treatment in a gravitational field in
~\cite{ful96},
where the gravitational induced effective mass 
can be calculated  from the mass shell condition, 
which is obtained by 
iterating the Dirac equation
\begin{equation}
(P_{\mu}+  P_{G\mu}  + A_{G\mu}{\cal P}_L)
(P^{\mu} +  P_{G}^{\mu}  + A_G^{\mu}{\cal P}_L) = m^2,
\end{equation}
where $P_{\m}$ is the 4-momentum, and  we have not included background
matter effects. However, the
following approximated conditions are usually valid for neutrinos, i.e.,
$P_{\mu} >> P_{G\mu}$,
$P_{\mu} >>  A_{G\mu}$ and  $P_{\mu} >>  m$.

The complication in calculating the neutrino phase 
in curved spacetime is related to the nature of the
neutrino trajectories. In flat spacetime, the neutrino
trajectories are straight lines. 
But in the curved spacetime, the geodesic is curved from the global  
point of view, which is more complicated than the situation of the flat
spacetime. 
Now we will  follow 
the factual mass particle world line to cope with the neutrino oscillation 
problem, but we can prove that the neutrino phase  along the null is the
half of the value along the geodesic in the high energy limit
(see APPENDIX A). However the relative phase of two mass 
neutrinos should be 
equivalent to that along the geodesic through dividing a factor of 2 
  because the arrival time difference of two mass neutrinos should be
considered\c{bha99}.  

In flat spacetime, the  phase factor can be written as a conventional
manner~\cite{ful96,ana,sto79},
\be
{\Phi} = \int m ds = \int (Edt -Pdx)
 = \int \eta_{\mu \nu}P^{\mu}dx^{\nu},
\label{phase}
\ee
where phase factor $\Phi $ is also the classical Langrange for the
particle motion, and the
route is along the particle world line(geodesic) determined by the
variational
principle or Jacobi-Hamilton equation~\cite{wud96,aud81,ana}.
{\em ds} is the interval in flat spacetime with
$ ds^2 =  \eta _{\mu\nu}dx^{\mu}dx^{\nu} $ and the metric  
$ \eta _{\mu \nu}=diag(+1, -1, -1, -1) $.
In the curved spacetime, however, the neutrino phase is calculated 
along the null\c{ful96,for96}  
\be
{\Phi} = \int m d\la =  \int g_{\mu \nu}P^{\mu}dx^{\nu} 
=\int P_{\mu} \sn{}^{\mu} d\la,
\label{phaseg}
\ee
where $\la$ is the affine parameter along the null.
 For the null (photon trajectory) it therefore may be convenient to
leave the
affine parameter {\em $\la$ } as the variable of integration in
Eq.(\ref{phaseg}). The tangent vector to the null 
${\stackrel{\circ}{n}}{}^{\mu} = dx^{\m}/d\la$,  and 
${x^{\m}}(\la) = \left[x^{0}(\la), x^{1}(\la), x^{2}(\la),
x^{3}(\la)\right]$. 
Now we consider the general mass shell condition and ignore the terms of
${\cal O}(A^2)$, ${\cal O}(AM)$,
${\cal O}(P_{G\mu}^2)$ and 
${\cal O}(P_{G\mu}M)$. Following ref.\c{ful96}, 
we obtain 

\be 
P_{\mu} \sn{}^{\mu} = m^2/2P_{o}  - A_{G\mu}\sn{}^{\mu} 
- P_{G\mu}\sn{}^{\mu}.
\label{pnu}
\ee

From Eq.(\ref{pnu}), we find that the total phase includes two types. 
The first type is contributed by the phase of the spinless particle  
$m^2/2P_{o}$, which is conventionally discussed \c{ful96,for96}
 and the second type is contributed by the ``spin connection".
According to the definition by Anandan~\cite{ana}, the type-I 
phase represents the contribution of the particle momentum coupled to 
the spacetime cuvature, and the type-II phase represents the spin
connection contribution  of the particle coupled to the spacetime
geometry. 

Further, it is convenient to define a column vector of flavor
amplitudes~\cite{ful96}. For example, for mixing between $\nu_e$ and
$\nu_{\tau}$,  

\be\label{xg}
\chi(\la) \equiv \pmatrix{\langle \nu_e | \Psi(\la)
                \rangle \cr
                \langle \nu_{\tau} | \Psi(\la) \rangle}.   
\ee
Eq.(\ref{xg}) can be written as a differential equation for
 the null  parameter $\la$,
\be
i {d\chi \over d\la} = \left(m^2 /2P_{o} - A_{G\mu}\sn{}^{\mu}
- P_{G\mu}\sn{}^{\mu}\right)
\chi,
\label{dcds}
\ee
Eq.(\ref{dcds})
can be integrated  to yield the
neutrino flavor evolution. A similar equation was obtained in
Ref. \cite{ful96}.

\section{Radial  neutrino oscillation in Schwarzschild
spacetime}

In this section we will study the phases along the geodesic and along
the null in Schwarzschild spacetime. 
 In Schwarzshild geometry, the spherically static spacetime can be
globally 
represented by the Schwarzshild coordinate system $\{x^{\mu}\}=
(t,r,\theta , \phi)$ with the line element as 
\be
ds^{2}= g_{oo}dt^{2} + g_{11}dt^{2}- r^{2}d\theta^{2} - r^{2}sin^{2}\theta
d\phi^{2}, 
\ee
\be
g_{oo} = (-g_{11})^{-1} = 1 - {r_{s}\over r },
\ee
where $r_{s} = 2 M $ is Schwarzschild radius  and M is the gravitational
mass of star. 
 The general form of tetrads in Schwarzshild spacetime are given as
follows (see APPENDIX B). 
The spacelike components of coordinates share the SO(3) rotational 
symmetry and timelike coordinate shares the reflection symmetry
\be
h^{a}{}_{\mu} \equiv \pmatrix{
\ga_{oo} & 0 & 0 & 0 \cr 
0 &\ga_{11}s\theta c \phi &r\;c \theta c \phi  &
-r\;s \theta s \phi \cr
0 &\ga_{11}s \theta s \phi &r\;c \theta s \phi  &
r\;s \theta c \phi \cr
0 &\ga_{11}c \theta  &- r\;s \theta   & 0} 
\ee
where $\ga_{oo}=\sqrt{g_{oo}} $, $\ga_{11}=\sqrt{-g_{11}} $, 
$s=\sin$ and $c=\cos$, and the subscript 
$\mu$ is the column index. This 
 tetrad  deduces the tetrad axial vector and tetrad  vector   
$a^{\mu} = 0$ and $v^{\mu}= (0, v^{1}, 0, 0)$. It is easy to understand
that 
the axial  tetrad is cancelled in the case of the spherical
symmetry, which represents the deviation from the spherical symmetry and
 occurs in the Kerr spacetime~\cite{nit81}. The nonzero tetrad
vector(see APPENDIX B)

\bea\nn
v_{1}& = &g_{11} v^{1}= - (g_{oo})^{-1}\left[ [ln\sqrt{g_{oo}}]^{'} + {1 -
\sqrt{g_{11}}\over r}\right] \\
     & = & {1\over2}(-g_{11})^{'} + {g_{11}\over r}(1 - \sqrt{-g_{11}}),
\eea   
so the type-II phase is

\bea \nn
\Phi_{II}& =& \int  P_{G\mu} dx^{\mu} = \int_{r_o}^r  v_{1}/2 dr\\\nn 
&=& {1\over 4}\left[ r -r_o + ln{(1-\rsr)\over (1- \rsro)} \right]\\ 
           &\simeq & {-1 \over 8} \left[ (\rsr)^{2} - (\rsro)^{2}
\right].
\label{p2}
\eea
From Eq.(\ref{p2}), if $r\rightarrow\rss$, the type-II phase will be 
infinite. Then in the weak field,  we
can conclude that 
 the type-II phase for neutrino in the curved spacetime
 is proportional to the square of the
gravitational potential, i.e., $(\rs2r)^2 $, which is
lower than the contribution  of the type-I phase that is related to the
gravitational potential, i.e., the gravitational red
shift~\cite{ful96,for96}. But, if we consider the strong gravitational 
field, e.g., near the horizon of the black hole\c{wud96}, the type-II 
phase might produce some effects on the particle interference. However,
 we stress that the type-II phase of Eq.(\ref{p2})  is
independent of the neutrino mass, namely it is equally contributed to the
different neutrinos. So this leads to the cancelling 
of the relative type-II phase between  the different mass neutrinos.
Therefore 
 we only consider the type-I phase of the mass neutrinos in the
Schwarzschild geometry by means of  the particle geodesic equation

\be 
{d^{2}x^{\lambda}\over ds^{2}} + {\stackrel{\circ}{\Gamma}}{} 
^{\lambda}_{\mu \nu}
{dx^{\mu}\over ds}{dx^{\nu}\over ds}=0
\ee
where
\be
{\stackrel{\circ}{\Gamma}}{}^{\lambda}{}_{\mu \nu} = \frac{1}{2}
g^{\lambda \rho} \left[ \partial_{\mu} g_{\rho \nu} + \partial_{\nu}
g_{\rho \mu} - \partial_{\rho} g_{\mu \nu} \right]
\label{lci}
\ee
is  the Levi--Civita connection. 
 Along
radial direction with $d\theta = d\phi = 0 $, there  are two 
nontrivial 
independent equations in the four geodesic equations(see Appendix C).
The nontrivial equation of timelike 

\be
\label{geq1}
\ddot{t} + {\dot{g_{oo}}\over g_{oo}}\dot{t} = 0
\ee
where the dot represents the derivative to {\em ds}, for instance,
$\dot{t}=
{dt \over ds}$. The Eq.(\ref{geq1}) gives

\be
\label{geq2}
g_{oo}\dot{t}= const.,~~~ or, ~~~E= P_{o} = m g_{oo}\dot{t}= const.
\ee
Eq.(\ref{geq2}) represents that the covariant energy is the motion
constant along the geodesic, by which the calculation of the phase is 
proceeded. It is stressed that it is the covariant energy $P_o$
(not $P^o$) the constant of motion, then this fact will be important 
in the later calculation of the neutrino phase. Otherwise the ambiguous 
definition of the neutrino energy in the gravitational field will lead to 
the confusion in understanding the gravitationally induced neutrino
oscillation.  

The mass shell condition can be written in the Schwarzshild spacetime,
which is not independent and can be deducted from the geodesic equations

\be
\label{mseq1}
g_{oo}\dot{t}^{2} + g_{11}\dot{r}^{2} + r^2 \dot{\phi}^2 = 1, ~~~
\ee   
substituting Eq.(\ref{geq2}) into Eq.(\ref{mseq1}) with $\dot{\phi} = 0$,
which means the radial motion, thus  we obtain

\be
\label{dsr1}
{ds \over dr}= \sqrt{{-g_{oo}g_{11} \over 
({P_{o}\over m})^{2} - g_{oo}}}.
\ee

If $m \rightarrow 0$, then ${ds\over dr}= 0$, the null condition is
recovered. If $r \rightarrow \infty$, the asympotic flat solution 
$g_{oo} \rightarrow 1$ and $-g_{11} \rightarrow 1$ is obtained, by which
we can acquire the phase of  flat spacetime. 
 In Schwarzshild spacetime, $-g_{oo}g_{11}= 1$, Eq.(\ref{dsr1})
becomes simply

\be   
\label{dsdr2}
{ds \over dr}= {1 \over \sqrt{ ({P_{o}\over m})^{2} - g_{oo}}}
= ({m\over P_{o}})(1 + \delta^2). 
\ee
where $\delta^2 = {g_{oo}\over 2} ({m \over P_{o}})^2$.
 The accurate phase factor in Schwarzshild spacetime along the radial 
direction is acquired in virtue of the  phase factor 

\bea
\nn
\Phi(geod)  & = & \int m({ds \over dr})dr
= \int {m dr \over \sqrt{ ({P_{o}\over m})^{2} - g_{oo}}}\\
\nn   & = &({mr\over k}) \sqrt{k + {r_{s}\over r}} - 
({mr_{o}\over k}) \sqrt{k + {r_{s}\over r_{o}}} \\\nn
&-&({mr_{s}\over 2 k^{3/2}})log\left[r_{s} + 2 kr + 2r\sqrt{k(k + {r_{s}\over
r})} \right]\\\nn   
&+&({mr_{s}\over 2 k^{3/2}})log\left[r_{s} + 2 kr_{o} + 2r_{o}\sqrt{k(k +
{r_{s}\over r_{o}})} \right]\\
      & \approx & { m^{2}\over P_{o}}(r - r_{o})+ O(\delta^{2}) =
2\Phi(null), 
\eea
where $k = ({P_{o}\over m})^{2} - 1 $.
$ \Phi(null) = { m^{2}\over 2P_{o}}(r - r_{o})$ is the neutrino phase 
along the null given by ~\cite{ful96,for96}. 

 It is noted that the
Schwarzschild coordinates r and $r_{o} $ are not the applicable physical 
distances, and the physical distance connected coordinate is given by 
$dL = \sqrt{-g_{11}}dr$, where dL is the physical distance. 
The quantity $0 \le g_{oo}\le 
1$ leads to  $ g_{oo}\ll k \sim 10^{12}$ for the electron neutrino. In
other words, the gravitational
contribution appears in one part of $10^{12}$ as shown in
Ref.~\cite{bha99}.
This means that the Schwarzschild
gravitation has little contribution to the neutrino oscillation phase,
which is  
same as that  concluded in ~\cite{ful96,bha99}. The  above phase
calculation is
a precisely  treatment, and we do not use any approxamations, 
except the final approximated simplification.
However, we admit that the Schwarzschild coordinates employed to calculate 
the phase factor is not the real physical distance although there exists 
some complications between them in the general case. 

In order to illustrate the validity of the  geodesic phase
treatment, we calculate the geodesic phase again from the other method. 
From the
equations (\ref{mseq1}) and (\ref{dsdr2}),
the derivative between the time-like and the space-like coordinates, 

\bea\label{dtdr}
{dt \over dr} &=& 
\sqrt{{(-g_{11})\over g_{oo}}}[1 - ({m \over P_{o}})^2 g_{oo}]^{-1/2}\\\nn
&\approx& \sqrt{{(-g_{11})\over g_{oo}}}(1 + \delta^2),\,\,\, 
\eea
If the neutrino mass $m \rightarrow 0$, then $\delta \rightarrow 0$, 
which leads to the null condition 
${dt \over dr} = \sqrt{{(-g_{11})\over g_{oo}}}$  
that is conventionally  used in the standard treatment\c{ful96,for96}. 
For the electron neutrino, the mass $m \sim 1 $ ev and the energy 
$E \sim 1$ Mev, $\delta ^2 \sim 10^{-12}$\c{boe92,bah94}, 
which is really a little quantity. 
If we set $\de = 0$, then Eq.(\ref{dtdr}) represents the null condition, 
which will produce the null phase~\cite{ful96,for96}. But, if this 
almost negligible quantity is taken into account, then what do  the 
things occur to the phase calculation? Let us follow the very routine 
phase calculation  as proceeded in the standard treatment of the neutrino 
phase in a gravitational field,

\be\label{cphi}
\Phi  =  \int g_{\mu \nu}P^{\mu} dx^{\nu}\\  = 
 \int (g_{oo} P^{o} dt + g_{11} P^{r} dr).
\ee 
The mass shell condition in Schwarzshild spacetime 

\be 
g_{\mu \nu}P^{\mu} P^{\nu} = m^{2},\,\,\,
g_{oo} (P^{o})^{2} + g_{11} (P^{r})^{2} = m^{2},
\label{mss}
\ee
and the approximated energy momentum relation from Eq.(\ref{mss})

\be \label{aep}
(-g_{11})P^{r}= \sqrt{g_{oo}(-g_{11})}P^{o} - \sqrt{{(-g_{11})\over
g_{oo}}}{m^2 \over 2P^{o}}, 
\ee
substituting (\ref{aep}) and (\ref{dtdr}) into (\ref{cphi}), we obtain

\bea
\label{rpap}
\nn
\Phi(geod)  & = & \int \left[ g_{oo} P^{o} {dt \over dr} + g_{11} P^{r}
\right]dr\\
\nn
      & = &\int \left[ \sqrt{{(-g_{11}) g_{oo}}} P^{o}\delta^{2} 
+ \sqrt{{(-g_{11}) g_{oo}}}{m^2 \over 2P_{o}} \right] dr\\
\nn
      & = &\int \left[ \sqrt{{(-g_{11}) g_{oo}}} {m^2 \over 2P_{o}}
+ \sqrt{{(-g_{11}) g_{oo}}}{m^2 \over 2P_{o}}\right] dr\\
      & = &\int \sqrt{{(-g_{11}) g_{oo}}} ({m^2 \over P_{o}})dr\\\nn
&=&{ m^{2}\over P_{o}}(r - r_{o})+ O(\delta^{2}) = 2\Phi(null). 
\eea
It is not strange that the geodesic  phase  is double of the result by
 the null~\cite{ful96}.  This also 
indicates that the application of the null condition arises the
correct result, and the geodesic 
phase  of two neutrinos despises the two
neutrino arrival time difference\c{bha99}. 

Now we discuss the physical coordinate problem because  the
Schwarzschild coordinate  employed in
Eq.(\ref{rpap}) is not the proper physical distance, however, which is
given by 

\bea
L - L_{o}
\null & \null \equiv \null & \null
\int_{r_o}^{r} \sqrt{-g_{11}}\,dr \\ \nn
\null & \null = \null & \null
r \sqrt{ 1 - \rsr}- r_o
\sqrt{ 1 - \rsro}\\ \nn 
&+& \rss \left[ \ln\!\left(\sqrt{ r - \rss }
+ \sqrt{r}\right) - \ln\! \left( \sqrt{ r_o - \rss }
+ \sqrt{r_o} \right) \right]\;.
\eea
We consider the case of a weak field, then the physical distance  is
approximately obtained 
\be
L - L_o \simeq r - r_o + {\rss \over 2} \ln \frac{r}{r_o}\;,
\label{apd}
\ee
where L($L_o $) is the proper physical distance corresponding to the 
coordinate r ($r_o $). In the case of weak field 
as well as  the conditions $\De L = L - L_o \ll L$ 
and $\De r = r - r_o \ll r$, we obtain

\be
\De r = ( 1 - {\rss \over 2L} )\De L = \left[ 1 + V(L)\right]\De L\;,
\label{rl}
\ee
where $V(L)= {- \rss \over 2L}$ is the gravitational potential expressed
by 
the physical distance. So the relative phase of the mass neutrinos is 

\be
\De \Phi ={ \De m^{2}\over 2P_{o}}\left[ 1 + V(L)\right]\De L\;,
\label{dpw}
\ee
where $\De m^2 = m_{\mu}^2 - m_e ^2$ is the mass square difference. So 
the oscillation length can be given by the usual manner

\be
L_{OSC} = 2\pi {2P_{o}\over \De m^{2}}\left[ 1 - V(L)\right]\;.
\ee
Likewise, as stated by Cardall and Fuller~\cite{ful96}, $P_o$ does {\em
not} 
represent the neutrino energy measured by a locally inertial observer at
rest at finite radius, but rather the energy of the neutrino
measured by such an observer
at rest at infinity. It is generally not possible to extract a
separate ``gravitational phase'' from the  expression (\ref{dpw});
nevertheless,
it is clear that gravity has an effect on the oscillations of the 
radially propagating neutrinos, corresponding to the
gravitational redshift. In the weak field limit one could
define a ``gravitational phase,'' however, the energy definition and
coordinate condition should be apparently declared.

\section{nonradial propagation in Schwarzshild spacetime}
In this section we contrast the radially propagating neutrinos
with the nonradically   propagating neutrinos in order to
demonstrate how the gravity affects vacuum neutrino oscillations.
As a further application, the general covariant phase can be applied to 
solve the nonrelativistic thermal neutron interference. 
 For the general nonradial propagation of the mass particle, the knowledge
of the geodesic should be exploited\c{mtw} (see APPENDIX C).  In the
$\theta = \pi/2$ plane. The
accurate phase along the azimuthal geodesic  can be calculated  by

\bea
\Phi^a (geod)  & = & \int m ds = \int m({ds \over dr})dr\\\nn
&=& \int {m dr \over \sqrt{ ({P_{o}\over m})^{2} - g_{oo}(1- \l2r2)}}\\
\nn   
& = &({mr\over k}) \sqrt{k + \rsr + \l2r2} -
({mr_{o}\over k}) \sqrt{k + \rsro + \l2r2}\\\nn
&-&({mr_{s}\over 2 k^{3/2}})log\left[r_{s} + 2 kr + 2r\sqrt{k(k +
\rsr + \l2r2)} \right]\\\nn
&+&({mr_{s}\over 2 k^{3/2}})log\left[r_{s} + 2 kr_{o} + 2r_{o}\sqrt{k(k +
\rsro + \l2r2)} \right]
\eea
where $l$ is the integral of motion $r^{2}\dot{\phi} = J/m = l
=const$, and J is the angular momentum of the particle. 
If $l = 0$, the radial geodesic phase of Eq.(\ref{rpap}) will be
recovered. 
 Equivalently, the nonradial phase can be calculated  in terms of the
azimuthal  coordinate $\phi$

\be
\Phi^a (geod) = \int m ds = \int m({ds \over d \phi}) d\phi = 
\int m(\dot{\phi})^{-1}  d\phi = \int {mr^2 \over l} d\phi 
\label{phasea}
\ee
then the
particle trajectory can be  determined by the trajactory
equation~\cite{mtw}
\be
{1\over r} = u = {r_s \over 2 l^2} (1 + e \cos \phi ),
\label{traj}
\ee
where e is the ellipse eccentricity, and 
we set the initial perihelion location at zero and 
neglect the perihelion precession term. Substituting Eq.(\ref{traj}) into 
Eq.(\ref{phasea}), we have, 
becomes 
\be
\Phi^a (geod) =  ({4 m l^3 \over r_s^2})I(\phi),
\label{phasea2}
\ee

\bea
I(\phi) &=&\int_{0}^{\phi} {d x\over ( 1 + e\cos x )^{2}}\\ \nn
     &=& -2 ArcTanh[{(e-1)Tan(\phi/2)\over \sqrt{e^2
-1}}](e^2-1)^{-3/2}\\\nn
    & +& {e \sin \phi \over (e^2 -1)(1 + e \cos \phi)}
\eea
For the low energy particle, the eccentricity of the trajectory can be
given 

\bea
e &=& \sqrt{1 + {2E J^2 \over m (r_s m/2)^{2}}}= |1 - 2 v^2
(\rrs)|,\\\nn  
2E/m &=& (v^2 - \rsr), ~~~ J = mvr. 
\eea
In the case of COW experiment, the scale of the instrument is about 5 
centimeters (shown in FIG.1.), which is much less than the radius of
the
earth, and so the transverse
azimuthal angle is very small $\phi= \de\phi$. Therefore, Taylor expansion
gives

\be
I(\de\phi)= {\de\phi \over (1 + e)^{2}}
= {\de\phi \over (1 + |1- 2v^2 \rrs|)^2}. 
\label{phitay}
\ee

\section{the covariant treatment of the  thermal neutron interference}

Now we use the  covariant phase to deal with the thermal neutron
interference on the earth (COW experiment), which is  satisfactorily 
 described by the COW interference phase obtained through solving 
Schoedinger equation 
 induced by the Newtonian 
gravitational potential\c{sak85}  or through the weak field approximation
method\c{sto79}. Our purpose of using the  covariant
method to cope  with the low energy thermal neutron interference is to find 
 how the spacetime global structure influences on the COW interference. Unlike the 
covariant treatment, the Newtonian
gravitational potential induced COW interference phase does not include 
 the information of the global structure of the curved spacetime, and
nonetheless the 
weak field approximation
method induced COW interference phase\c{sto79} may also neglect some 
characteristics 
of the curved spacetime. The second motivation of  using  the  covariant
 method  to deal with the thermal neutron
interference aims at indicating that the  covariant phase is not only useful 
 for the high energy mass neutrinos but also successfully 
applicable to the low energy thermal neutron interference.  
The particle wave behaviours incorporated into the curved spacetime will 
reflect some quantum properties of the particle in 
Einstein's general relativity, 
 which has not yet  been well explored until now. 
 Moreover the successfully application of the  covariant phase 
to the COW experiement  can help us to wipe out
the
suspect of the universality of the covariant treatment of the phase on the
particle interference in any energy bands.

In FIG.1., 
the scale of the instrument of COW experiment is some centimeters, which
is much less than the scale of the gravitational source, the radius of
the earth. 
 For the arm plane of the COW instrument perpendicular to the surface of
the earth, the
thermal neutron motions are divided into two beams, the one is radial
motion along AB
 (CD) and the other is nonradial along AC (BD). We calculate the phase
difference along the different routes ABD and ACD. Taking the
earth as
 an exact sphere, the Schwarzschild metric induced radial velocity is
\be
{P \over m}={dr \over ds} = \sqrt{(E/m)^{2} - g_{oo}} = \sqrt{k +
\rsr}\;,
\ee

so the momentum difference from A to B

\bea
(P_{A}^{2} - P_{B}^{2})/m^2  &=& {r_{s} \over R_{\op}}- {r_{s} \over    
R_{\op} + H} \\\nn
&=& 2g H  + O(\de_s \de_{H}^2),
\eea
where $\de_{s} = {r_{s} \over R_{\op}} \sim 10^{-11}$ and
 $\de_{H} = {H \over R_{\op}}\sim 10^{-8}$,
$g = {r_{s} \over 2R_{\op}^2}$ is the
gravitational acceleration on the surface of the earth and $R_{\op}$ is
the earth radius.
 
For the nonradial motion of the  nonrelativistic thermal neutron beams through
AC and BD, with  the velocity  condition $2v^2 \rrs > 1$ (in the
COW
experiement, $v^2 \sim 10^{-10}$\c{cow} and $\rsr\sim 10^{-11}$), from 
 Eq.(\ref{phitay}), we have 

\be
I(\de\phi)={\de\phi \over (1 + |1- (v/v_G )^2 |)^2} 
 =   ({v_{G} \over v} )^{4}\de\phi , 
\label{idp2}
\ee
where $v_G = \sqrt{\rs2r}$ 
and the velocity approximation condition now  becomes ${v\over v_G} > 1
$, which 
yields the transverse phase at radius r,  from Eq.(\ref{phasea2})  
   and Eq.(\ref{idp2})
\be
\label{nrpww}
\Phi^{a}(r) = ({mr v^3 \over v_G^4})I(\de \phi) = {m^2 r \de\phi \over
P},~~~
\ee
where $P = mv$ is the transverse momentum, and the Schwarzschild
coordinate r should be replaced by the physical
distance if applicable, {\it i.e.}, $dL = \sqrt{-g_{11}} dr$, where L
notes the physical distance.  In the case of the weak field, 
the relation between the coordinate difference and the physical distance 
is 
$\De r = (1 + {\rss \over 2L})\De L$. Thus Eq.(\ref{nrpww})
 becomes 
\be
\Phi(L) = {m^2 H \over P}(1 + {\rss \over 2L}),
\ee
where H is the arm length of the the COW interference instrument
(FIG.1),  and we set the square route, i.e., H=AB=BD=AC=CD in FIG.1.
Then
the interference phase difference of two thermal neutron beams becomes

\bea
\De\Phi(COW) &=& \Phi(ABD) - \Phi(ACD)\\\nn
& =&\Phi(R_{\oplus}+ H) - \Phi(R_{\oplus})\\\nn
&=&{m^2 H\over \lan P\ran}g H = m^2 g H^2 {\la_{n} \over 2\pi}\;.
\label{nrwdp}
\eea
where $\la_{n} =  2\pi /\lan P\ran$ is the thermal neutron wave length 
with the momentum $\lan P\ran = (P_{A} + P_{B})/2 \sim P_{A}\sim P_{B} $.
This is just
the anticipated result  in COW experiment~\cite{cow}

However, we stress two facts here, from the point of view of the covariant
treatment of Dirac particle. Firstly, the COW 
thermal neutron phase obtained is a 
type-I phase because  the type-II phase is the square of the gravitational
redshift, which can be neglected in the weak field condition. 
Secondly, 
 if the thermal velocity satisfies $v < v_G$, unlike the COW phase, the
interference will follow a
different regulation related to the thermal neutron wave length. The 
above condition cannot be obtained from the weak field expansion method  
to deal with the COW neutron interference\c{sto79}.  
 In other words, we can construct the critical wave length for COW 
thermal neutron  as 
$\lambda_G = { 2\pi \over mv_G}$, therefore  the 
critical condition for the COW interference phase needs the
 wave length of the thermal neutron  must be shorter than this critical
wave length  
, i.e., $\lambda_n < \lambda_G $.

\section{conclusions and discussions}

On the basis of the particle interference phase along the geodesic, the
following conclusions are obtained and demonstrated. 

(1) On the phase of the particle propagating in Schwarzschild spacetime,
we exploit the geodesic  route to calculate  it, and 
our results for the neutrino oscillation relative phase  is in agreement
with that in Refs.~\cite{ful96,for96} if the geodesic phase is
divided by a factor of 2, which originates from the consideration of 
 the  simultaneous arrival of two mass 
neutrinos for the interference\c{bha99}. However  our
results are different from that in ref.~\cite{ahl96}, where the weak
field expansion is applied. We find that two factors influence on our 
understanding of the gravitationally induced neutrino oscillation,  
 i.e., energy definition and physical distance.  The
geodesic phase will produce a factor of 2 error contrasting to the null 
phase, which
is also  
paid attention in the flat spacetime\c{lip95}, and then the energy
definition 
needs the integral constant of motion in the gravitational field.
Moreover, the physical distance
should be transformed from the Schwarzschild coordinate system. 
Otherwise any
ambiguous
definitions of above two conditions will result in confusion or 
conflict because the mass neutrino phase is very sensitive to these
conditions.

(2) Two types of phases are taken into account in the article. Type-I 
phase is contributed by the curved spacetime, which represents the
coupling 
of the momentum of the particle to the curved spacetime. 
 Type-II phase is contributed by the coupling
of the ``spin connection" of the particle to the curved spacetime. In the 
case of Schwarzschild weak field, type-II phase is proportional to the 
square of the effect of gravitational red shift. Namely, type-I phase 
is the first order effect of the gravitational potential and type-II phase
is the second order effect of the gravitational potential. So we can 
neglect the type-II phase when dealing with the physical problems in the
solar 
system, such as atomspheric neutrino, solar neutrino as well as the
thermal neutron interference  in COW experiment.

(3) We find that, if  the thermal neutron
velocity 
exceeds over a critical velocity,   then COW 
interference phase will appear. But, if the thermal neutron velocity is
 not higher than this velocity, the COW phase will be broken. This
critical 
velocity may be the critical condition for the validation of the COW
interference\c{cow}, however this conclusion cannot be derived  from 
the weak field method to deduce the COW interference phase\c{sto79}. 
 From the quantum mechanics language, there exists a critical wave 
length, and
the 
COW interference needs that the thermal neutron wave length must be 
 shorter  than this critical wave length, $\lambda_n < \lambda_G $ 
with $\lambda_G = { 2\pi \over mv_G}$. 
 If this conclusion is valid, 
we could suggest the low velocity thermal
neutron interference 
experiement to inspect the proposed effect. Moreover 
the   further experiement not only detects the Newtonian gravitational 
potential induced interference effect, but also becomes a sensor of 
the curved spacetime, to inspect the  general relativity 
induced COW interference condition.

In summary, we can speculate that the type-II phase should play an
important role in the strong rotational gravitational field, especially
near 
Kerr black hole, where the high order terms of gravitational potential 
will effect.

\section*{appendix A}

The velocity of an extremely relativistic neutrino is nearly the speed of
light. Despite of this, the propagation difference between a massive
neutrino
and a photon can have important consequences. In the standard treatment of
the neutrino oscillation, the neutrino is usually supposed to travel along
null-lines~\cite{zub98,bil98,boe92,bah94}, and almost no attention
has
been paid to this small difference. Although seemingly irrelevant, this
tiny
deviation becomes  important for the understanding of the neutrino
oscillation.  Motivated by this argument, we will compare the
neutrino
phase when calculated along the geodesic and along the null-line. With
this,
we will be able to verify the factor of 2  error  when the null 
is replaced by the geodesic. This study can be shown to remain valid in
the case of flat spacetime.

Let $n^{\mu}$ and ${\stackrel{\circ}{n}}{}^{\mu}$ be the tangent vectors
to
the geodesic and to the null-line, respectively, their difference
$\ep^{\mu}$
being a small quantity for the case of an extremely relativistic neutrino.   
Here, we suppose that the two neutrinos, the massless and massive, start  
their journey at the same initial spacetime position A, and their {\it
space}
routes are almost the same. But, there will be an arrival time-difference
at
the  detector position B. This means that their 4-dimensional spacetime
trajectories are not the same, and consequently the tangent vectors will
present a small difference. Thus, we have
\be \label{nne}
n^{\nu} = {\stackrel{\circ}{n}}{}^{\nu} + \ep^{\nu}, \quad \mbox{or} \quad
P^{\nu} = {\stackrel{\circ}{P}}{}^{\nu} + m\ep^{\nu} \; ,
\ee
where $P^{\nu} = m n^{\nu}$ (${\stackrel{\circ}{P}}{}^{\nu}= m
{\stackrel{\circ}{n}}{}^{\nu}$) is the 4-momentum along the geodesic
(null-line) with
\be
{\stackrel{\circ}{n}}{}^{\nu} = {dx^{\mu}\over d\lambda} =
{d{\stackrel{\circ}{x}}{}^{\nu}\over ds}
\ee
and
\be
n^{\mu} = {dx^{\mu}\over ds} \; .
\ee
In these expressions, $\lambda$ and $s$ are respectively affine parameters
along the null and the geodesic lines. These two tangent vectors satisfy
the
mass shell relations of the geodesic and the null-line:
\be\label{gn}
g_{\mu\nu} n^{\mu}n^{\nu} = 1
\ee
and
\be\label{nn}
g_{\mu\nu} {\stackrel{\circ}{n}}{}^{\mu}{\stackrel{\circ}{n}}{}^{\nu} = 0
\; .
\ee
Now, substituting (\ref{nne}) into (\ref{gn}), we obtain
\be \label{msg2}
g_{\mu\nu} ({\stackrel{\circ}{n}}{}^{\mu} + \ep^{\mu})   
 ({\stackrel{\circ}{n}}{}^{\nu} + \ep^{\nu}) = 1 \; ,
\ee
or, by using (\ref{nn}),
\be \label{nne2}
2 g_{\mu\nu} {\stackrel{\circ}{n}}{}^{\mu} \ep^{\nu} +
{\cal O}(\ep^{2}) = 1 \; .
\ee
We can estimate the order of $\{n^{\mu}\}$ and $\{
{\stackrel{\circ}{n}}{}^{\mu} \}$ by noting that  
$ n \ep \sim 1/2 $, which implies that $\ep \sim n^{-1} \sim {m\over E}$,
where $E = P^{o} \sim P^{i}$ {} $(i=1,2,3)$ for a relativistic neutrino.
   
The neutrino phase induced by the null condition, as in the standard
treatment, comes from the 4-momentum $P^{\nu}$ defined along the geodesic   
line, and the tangent vector $\{ {\stackrel{\circ}{n}}{}^{\mu} \}$ to the
null-line~\cite{ful96}. We notice that, if the 4-momentum
${\stackrel{\circ}{P}}{}^{\nu}$ defined along the null-line was instead
used
to compute the null phase, we would obtain zero because of the null
condition.
Therefore, the  phase along the geodesic line (geodesic phase) and the
phase
along the null-line (null phase) can be written respectively
as~\cite{ful96,aud81,sto79,ana}
\be\label{nrpb}
\Phi({\rm geod})= \int m ds = \int
g_{\mu\nu}P^{\mu}n^{\nu}ds \; ,
\ee
and
\be
\Phi({\rm null}) = \int g_{\mu\nu}P^{\mu}{\stackrel{\circ}{n}}{}^{\nu}ds
\; .
\ee
Therefore, the difference between the geodesic phase and the null phase,
by using Eq.(\ref{nne2}), is
\bea \nn
\Phi({\rm geod})& -& \Phi({\rm null})= \int
g_{\mu\nu}P^{\mu}(n^{\nu} - {\stackrel{\circ}{n}}{}^{\nu}) ds\\\nn
&= &\int g_{\mu\nu}P^{\mu} \ep^{\nu} ds  = 
 \int g_{\mu\nu}{\stackrel{\circ}{P}}{}^{\mu}\ep^{\nu} ds +
{\cal O}(\ep^{2})\\\nn
&=& {1\over2}\int m ds + {\cal O}(\ep^{2})
 =  {1\over2}\Phi({\rm geod}) + {\cal O}(\ep^{2}) \; ,
\eea
that is
\be
\Phi({\rm geod}) = 2 \Phi({\rm null}) + {\cal O}(\ep^{2}) \; .
\ee
This conclusion, valid for a general curved spacetime, is similar to that
found in in a Schwarzschild~\cite{bha99} spacetime. Concerning
the Schwarzschild spacetime, Bhattacharya {\it et al}~\cite{bha99} have
the
following argument for the factor of 2. As the neutrino energy is fixed,
but the masses are different, if a interference is to be observed at the
same
final spacetime point B$(r_B,t_B)$, the relevant components of the wave
function could not both have started at the same initial spacetime point
 A$(r_A,t_A)$ in the semiclassical approximation. Instead, the lighter
mass   
(hence faster moving) component must either have started at the same time
from
a spatial location $r<r_A$, or (what is equivalent) started from the same
location $r_A$ at a later time $t_A + \Delta t$. Hence, there is already
an
initial phase difference between the two mass components due to this time
gap,
even before the transport from $r_A$ to $r_B$ which leads to the phase
$\Phi({\rm null})$, {\it i.e.}, the additional initial phase difference
may be
taken into account~\cite{bha99}.

\section*{appendix B}
Tetrad in the spherical, static and isotropic coordinate system 
$(X^{0}=t, X^{1}, X^{2}, X^{3})$, $h^{(0)}{}_{0}= \sqrt{C(\rho)}$  and 
$h^{a}{}_{\al}= \sqrt{D(\rho)}\delta^{a}_{\al}$ \c{hay79}, the
corresponding 
line element (to avoid confusion, Latin indices are enclosed in 
parentheses to express Lorentz coordinates)
\bea
ds^{2}&=& g_{\mu\nu}dX^{\mu}dX^{\nu}\\\nn
&= & C(\rho)dt^{2}- D(\rho)(d\rho^{2}
+ 
\rho^{2}d\Omega^{2})
\eea

\bea
d\Omega^{2}= d\theta^{2} + \sin\theta d\phi^{2},~~&\,\,\, 
X^{1}= \rho \sin\theta \cos\phi\\
X^{2}= \rho \sin\theta \sin\phi,~~&\,\,X^{3}= \rho \cos\theta 
\eea 

comparing the line element in the Schwarzschild form, we get
\bea 
C(\rho)= g_{oo},~~&\,\,\sqrt{D(\rho)} \rho = r, ~~&~~ 
{\partial \rho \over \partial r } = \sqrt{{-g_{11}\over D(\rho)}}
\eea
using the general coordinate transformation
\be
h^{a}{}_{\mu} = {\partial X^{'\nu} \over \partial  X^{\mu} }
h^{'a}{}_{\nu}
\ee

where $\{X^{\mu}\}$ and $\{ X^{'\nu}\}$ are the isotropic and 
Schwarzschild 
coordinates respectively. Through using the above transformation and 
relationship, then we obtain the tetrad in the Schwarzschild coordinate 
system.  
\be
\label{te1}
h^{a}{}_{\mu} \equiv \pmatrix{
\ga_{oo} & 0 & 0 & 0 \cr 
0 &\ga_{11}s\theta c \phi &r\;c \theta c \phi  &
-r\;s \theta s \phi \cr
0 &\ga_{11}s \theta s \phi &r\;c \theta s \phi  &
r\;s \theta c \phi \cr
0 &\ga_{11}c \theta  &- r\;s \theta   & 0}, 
\ee
where $\ga_{oo}=\sqrt{g_{oo}} $, $\ga_{11}=\sqrt{-g_{11}} $, 
$s=\sin$ and $c=\cos$, and the subscript 
$\mu$ is the column index. 
The contravariant tetrad can be constructed by $h_{a}{}^{\mu} = g^{\mu\nu}
h^{b}{}_{\nu}\eta_{ab}$, 

\be
\label{te2}
h^{a}{}_{\mu} \equiv \pmatrix{
\ga_{oo}^{-1} & 0 & 0 & 0 \cr 
0 &\ga_{11}^{-1}s\theta c \phi &r^{-1}\;c \theta c \phi  &
-(r\;s \theta)^{-1} s \phi \cr
0 &\ga_{11}^{-1}s \theta s \phi &r^{-1}\;c \theta s \phi  &
(r\;s \theta)^{-1} c \phi \cr
0 &\ga_{11}^{-1}c \theta  &- r^{-1}\;s \theta   & 0}. 
\ee
One can proof that the orthagonal 
relation between $h^{a}{}_{\mu}$ and $h^{a}{}_{\mu}$ are satisfied. 
Eqs.(\ref{te1}) and (\ref{te2}) 
can construct the  tetrad vector and the tetrad axial-vector 

\bea
\Ga^{o}{}_{o1} = [ln \sqrt{g_{oo}}]^{'}, \,\,\Ga^{3}{}_{32}=
\Ga^{3}{}_{23} = \tan \theta\\\nn
\Ga^{1}{}_{11}= [ln \sqrt{-g_{11}}]^{'}, \,\,\,
\Ga^{1}{}_{22} = {r \over \sqrt{-g_{11}}}\\\nn
\Ga^{1}{}_{33}=\Ga^{1}{}_{22}
(\sin\theta)^{2},~~~ \,\, \Ga^{2}{}_{33}= - \sin \theta
\cos\theta\\\nn
\Ga^{2}{}_{21} = \Ga^{3}{}_{31}={1 \over r},~~ \,\, \Ga^{2}{}_{12}
= \Ga^{3}{}_{31}={\sqrt{-g_{11}}\over r}\\\nn
\sqrt{-g_{11}}^{'} = {d \sqrt{-g_{11}} \over dr}
\eea

\bea
T^{0}{}_{01} = [ln \sqrt{g_{oo}}]^{'},~~\,\,
h = r^{2}\sin\theta,\\\nn 
T^{2}{}_{21} = -T^{2}{}_{21}  = (1 - \sqrt{-g_{11}})/r,
\eea

\bea
a^{\mu}&\equiv& 0\\ \nn
v^{1} & = & \left[ (ln\sqrt{g_{oo}})^{'} + 
{1 - \sqrt{-g_{11}}\over  r}\right] 
\approx {1\over 2r}({r_{s}\over r})^{2}, 
\eea
\bea
v_{1} & = & g_{11} v^{1}= - (g_{oo})^{-1}\left[ [ln\sqrt{g_{oo}}]^{'} 
+ {1 -\sqrt{g_{11}}\over r}\right]\\ \nn
& = & {1\over2}(-g_{11})^{'} + {g_{11}\over r}(1 - \sqrt{-g_{11}})
\eea

\section*{appendix C}
In the Schwarzschild geometry, the components of the geodesic equation 
are\c{mtw}

\bea
\ddot{t} + {\dot{g_{oo}}\over g_{oo}}\dot{t}& =& 0\\ \label{geqapp1} 
\ddot{r} + {1\over 2}g_{oo}^{'}(-g^{11}) (\dot{t})^{2}&+& \\\nn
ln\sqrt{-g^{11}}^{'} - r g^{11}[ (\dot{\theta})^{2} + 
 \sin^{2}\theta (\dot{\phi})^{2}]& =& 0\\ \label{geqapp2}
\ddot{\theta} + {2 \over r}\dot{\theta}\dot{r} - 
\sin\theta \cos \theta(\dot{\phi})^{2}& =& 0\\ \label{geqapp3}
\ddot{\phi} + {2 \over r}\dot{\phi}\dot{r} + 2   
\cot\theta \dot{\theta}\dot{\phi}& =& 0\label{geqapp4}
\eea

where the dots indicate  derivatives with respect to s, such as 
 $\dot{t}\equiv {dt \over ds}$ and $\dot{r}\equiv {dr \over ds}$. 
We will assume that the orbit is in the plane $\theta = {\pi \over
2}$,
then Eqs.(\ref{geqapp1}) and (\ref{geqapp4}) can now be integrated
directly, giving two integrals of the motion

\bea
g_{oo}\dot{t} = {E \over m}=const,\\\nn
E= P_{o},
~~r^{2} \dot{\phi} = {J \over m}= l = const,~~~J=-P_{3}
\eea
 These two constants are proportional 
to the energy and the angular momentum, respectively. Precisely, these 
quantities are constants of the motion and represent the energy and the
angular momentum per unit mass.

\begin{figure}
\center{
\leavevmode\epsfysize=5.4cm \epsfbox{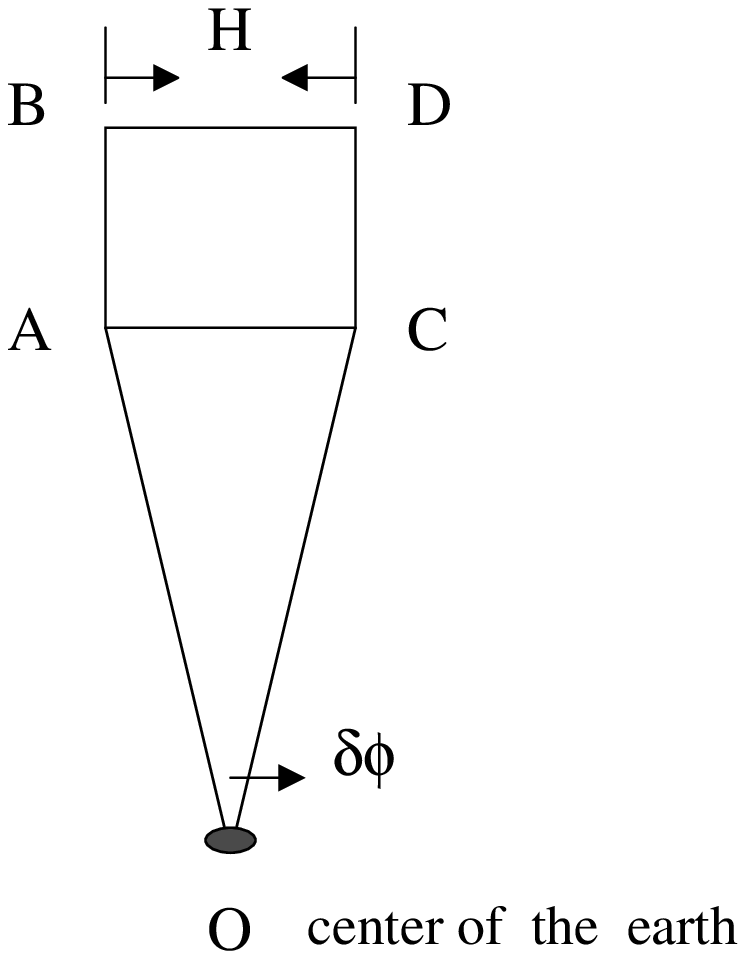} }
\vspace{10pt}
{\caption{\small
 Diagram illustration of the thermal neutron interference route.
We assume the arms AB, BD, AC and CD  to be same, i.e., AB = BD = AC =
CD = H. OA and OC represent
the radius of the earth and the angle between OA and  OC is $\de \phi$. }}
\label{fig2}
\end{figure}

\end{document}